\documentstyle[12pt]{article}
\begin{document}
\large
\newcommand{\be}{\begin{equation}}
\newcommand{\ee}{\end{equation}}
\newcommand{\bea}{\begin{eqnarray}}
\newcommand{\eea}{\end{eqnarray}}
\newcommand{\bi}{\begin{itemize}}
\newcommand{\ei}{\end{itemize}}
\newcommand{\bc}{\begin{center}}
\newcommand{\ec}{\end{center}}

\title {A closed form for Consistent Anomalies in Gauge Theories}
\author{Silvio P. Sorella \\
and\\
Liviu T\u{a}taru\thanks{Supported by \"{O}sterreichisches Bundesministerium
f\"{u}r Wissenschafl und Forschung}
\thanks{Permanent address:Dept.Theor.Physics, University of Cluj,
Romania.}\\[3.0cm]
Institut f\"{u}r Theoretische Physik\\
Technische Universit\"{a}t Wien\\
Wiedner Hauptsra\ss e 8-10\\
A-1040 Wien (Austria)   }
\maketitle

\vspace{4mm}
\centerline{ \bf REF. TUW 93-15}

\vspace{4mm}
\begin{abstract}
The new method for solving the descent equations for gauge theories
proposed in \cite{s} is shown to be equivalent with that based on
the {\em "Russian formula"}. Moreover it allows to obtain in a closed
form the expressions of the consistent anomalies
in any space-time dimension.
\end{abstract}

\setcounter{page}{0}
\thispagestyle{empty}

\newpage
\section{Introduction}
Consistent anomalies in gauge field theories occur
when the gauge invariance cannot be maintained at the
quantum level.

In this case the variation
 of the connected vacuum functional in the presence
 of external gauge fields
 does not vanish. This implies that the gauge
currents are no longer covariantly conserved but
have the anomalies as their divergence. As a direct consequence
of its definition the anomaly must satisfy the well
known Wess-Zumino  consistency
conditions \cite{wz}.
These conditions, when formulated in terms
 of the BRST transformations \cite{brs,t},
yield a cohomology problem for the nilpotent BRST operator $s$
\be
s \Delta=0 \label{eq:an}
\ee
where $\Delta$ is the integral of a local polynomial
in the fields and their derivatives. An  useful and powerful
method to find the non-trivial solutions of Eq.(1) is
given by the {\em descent-equations}
technique\cite{z,zwz,bz,z2,msz,dtv,baulieu,bonora,cotta,bdk}.

Writing $\Delta=\int{\cal A}$ Eq.(1) translates into
the local condition
\be
s{\cal A}+d{\cal Q}=0    \label{eq:d}
\ee
where $d$ is the exterior differential on the flat
space-time M, ${\cal Q}$ is a local polynomial in the fields and
\be
s^2=d^2=s d+d s=0
\ee
It is easy to check that Eq.~(\ref{eq:d}), due to the vanishing of the
cohomology of $d$, generates a tower of descent equations:
\bea
s{\cal A}+d{\cal Q}=0 \nonumber \\
s{\cal Q}+d{\cal Q}^1=0 \nonumber\\  \label{eq:de}
\cdots \nonumber\\
\cdots \nonumber\\
s{\cal Q}^{k-1}+d{\cal Q}^k=0 \nonumber\\
s{\cal Q}^k=0
\eea
with ${\cal Q}^i$ local polynomials in the fields.

These equations, as it is well
known since several years, can be solved by means of a transgression
procedure generated by the so called
{\em "Russian formula"}\cite{z,msz,dtv,baulieu,stora}.

More recently a new way
of finding nontrivial solutions of the
tower (4) has been proposed by one of the authors
and successfully applied to the study of the Yang-Mills
cohomology \cite{s} and of the gravitational anomalies \cite{ws}

 The method is based on the introduction of an operator $\delta$
which allows  to
express the exterior derivative $d$ as a BRST commutator
\be
d=-\left[ s ,\delta \right]    \label{eq:so}
\ee
It is easy to prove that, once the decomposition (5) has been found,
repeated applications of the operator $\delta$ on the polynomial
${\cal Q}^k$ which solves the last equation of Eqs.(4) give an explicit
solution for the other cocycles ${\cal Q}^i$ and for the anomaly ${\cal A}$.

It is interesting to observe that the
decomposition (5) occurs also in topological field theories \cite{olivier}
and in the bosonic string \cite{wss} . For these models the operator
$\delta$ is the generator of the topological vector supersymmetry and
allows for a complete classification of anomalies and counterterms.

Let us emphasize that solving the last equation of
the tower (4) is a problem of local cohomology instead of
a modulo-$d$ one. It is apparent that, thanks to the operator
 $\delta$, the characterization of the cohomology of $s$ modulo $d$
is essentially reduced to a problem of local cohomology.
Let us recall also that the latter has been solved in
the case of the gauge theories \cite{dtv,baulieu,bdk,dhtv}
and, more generally, it can be systematically
studied by using several methods as, for instance, the spectral
sequencies technique \cite{d,b,bl}. We have thus an alternative tool
for an algebraic characterization of the cohomology
of $s$ modulo $d$. It remains, however, to prove the complete equivalence
of the expressions obtained by the decomposition (5) with
the ones given by the {\em "Russian formula"}. This is the aim of
this paper, to give an explicit proof of the equivalence
of the two methods.

Actualy, as we shall see in details, the proof of the equivalence
turns out to be a direct consequence of the vanishing of the cohomology
of $(d+s)$ and of an elegant and quite simple formula
which allows to collect into a unique closed equation the solution
of the tower ~(\ref{eq:de})  for any space-time dimension and ghost number.
In the following we will take as explicit example the case of the
Yang-Mills gauge theories, the result being easily extended
to the gravitational anomalies and to the topological
models.

The paper is organized as follows. In Sect.2 we introduce the
differentials $d$ and $s$ and we briefly recall
some properties concerning their cohomologies. In Sect.3 we
introduce the operator $\delta$ and we prove the equivalence with
the {\em "Russian formula"}. In Sect.4 we
present a simple closed expression for the solution of the descent
equations. Finally, Sect.5 is devoted to some examples.

\section{Notations and Conventions}

To fix the notations let us introduce the local space ${\cal V}$
of form-polynomials \cite{dtv} in the variables $(A, dA, c, dc )$;
$A$ and $c$ being respectively the one-form gauge
connection $ A=A_{\mu}dx^{\mu} $ and the zero form ghost field. All the
fields are Lie algebra valued; $  A_{\mu}=T_aA^a_{\mu}$ and $c=T_ac^a$
with $\{ T_a\}$ the antihermitian generators of a semisimple Lie group $G$
in some finite representation. The BRST transformations are
\bea
sA&=&Dc=dc+\left[ A,c\right], \nonumber\\
sc&=&c^2=\frac{1}{2}\left[c,c\right].
\label{eq:brst}
\eea
where $\left[a,b\right]=a b-(-1)^{\mid a\mid \mid b\mid}b a$
denotes the graded commutator and $\mid a\mid$ is the degree of $a$ defined
as the sum of its ghost number and of
 its form degree. $A$ and $c$ have both degree one.

The two-form field strength $F$ reads
\be
F=dA+A^2      \label{eq:f}
\ee
and
\be
dF=\left[ F ,A \right] \label{eq:bi}
\ee
is its Bianchi identity.

To study the cohomology of $ s$ and $d$ on the space of form-polynomials
we switch, following \cite{dtv}, from $(A, dA, c, dc)$ to the
more convenient set of variables  $(A,F,c,\xi=dc)$, i.e. we replace
everywhere $dA$ with $F$ by using Eq.~(\ref{eq:f}) and we introduce
the variable $\xi=dc$ to emphasize the local character of
descent equations~(\ref{eq:de}). On the local space
${\cal V}(A, F, c, \xi)$
the action of the differentials $ s $ and $d$ is given by
\bea
sA=\xi+\left[c,A \right] \hspace{1.2cm}&,&\hspace{1.2cm} sF=\left[ c,F\right];
 \nonumber \\     \label{eq:s}
sc=c^2 \hspace{1.2cm} &,&\hspace{1.2cm}s\xi=\left[ c,\xi \right]
\eea
and
\bea
dA=F-A^2 \hspace{1.2cm}&,&\hspace{1.2cm}
 dF=\left[F,A\right] \nonumber \\  \label{eq:dd}
 dc=\xi \hspace{1.2cm}&,&\hspace{1.2cm} d\xi=0
\eea
In particular, from Equations~(\ref{eq:s}) and~(\ref{eq:dd})
it follows that on ${\cal V}(A, F, c, \xi)$ both $ d $ and
$ (d+s)$ have vanishing cohomology \cite{dtv,cotta,sull}.

This is  easily proved by introducing the counting (filtering) operator
${\cal N}$
\bea
{\cal N}A=A \hspace{2cm}&,&\hspace{2cm} {\cal N}F=F \nonumber\\
{\cal N}c=2c\hspace{2cm}&,&\hspace{2cm} {\cal N}\xi=2\xi
\eea
according to which the exterior derivatives $d$ and $\hat{d}=(d+s)$
decompose as
\bea
d&=&d^{(0)}+d^{(1)}, \nonumber\\
\left[{\cal N}, d^{(\nu)}\right]&=&\nu d^{(\nu)}\hspace{1.3cm}\nu=0,1
\eea
and
\bea
\hat{d}&=&\hat{d}^{(0)}+\hat{d}^{(1)}+\hat{d}^{(2)},        \nonumber\\
\left[ {\cal N} ,\hat{d}^{(\nu)} \right]&=&\nu
                 \hat{d}^{(\nu)}\hspace{1cm}\nu=0,1,2
\eea
with
\bea
d^{(0)}A &=& \hat{d}^{(0)}A=F \nonumber\\
d^{(0)}c &=&\hat{d}^{(0)}c=\xi     \label{eq:do}
\eea
and
\be
d^{(0)} d^{(0)}=0 \hspace{2cm} \hat{d}^{(0)} \hat{d}^{(0)}=0
\ee
{}From Eqs.~(\ref{eq:do}) it is apparent that $d^{(0)}$ and $\hat{d}^{(0)}$
have vanishing cohomology, it then follows that also $d$ and $(d+s)$
have trivial cohomology due to the fact that the cohomology of $d$
(resp.$(d+s)$) is isomorphic to a subspace of the cohomology
of $d^{(0)}$ (resp.$(d+s)^{(0)}$)\cite{d,b,bl}.
For what concerns the cohomology of $s$ it turns
out \cite{dtv,baulieu,cotta,bdk,stora,dhtv}
that it is spanned by polynomials in the variables $(c , F)$
generated by elements of the form
\be
 \left( Tr\frac{c^{2m+1}}{(2m+1)!} \right) \cdot {\cal P}_{2n+2}(F)
\hspace{1.3cm} m,n=1,2,\cdots  \label{eq:gl}
\ee
with ${\cal P}_{2n+2}(F)$ the invariant monomial of degree
$(2n+2)$:
\be
{\cal P}_{2n+2}(F)=TrF^{n+1}.
\ee
The Bianchi identity ~(\ref{eq:bi})  implies that $ {\cal P}_{2n+2}(F)$
is $d$-closed
\be
{\cal P}_{2n+2}(F)=d\omega^0_{2n+1} \label{eq:tf1} .
\ee
Using the triviality of the cohomology of $d$ one immediately
gets the tower of descent equations:
\bea
s\omega^0_{2n+1} + d\omega^1_{2n} &=& 0 \nonumber\\
s\omega^1_{2n} + d\omega^2_{2n-1} &=& 0 \nonumber\\
&\cdots&  \nonumber\\
&\cdots&   \nonumber\\     \label{eq:tow}
s\omega^{2n}_1 + d\omega^{2n+1}_0 &=& 0 \nonumber\\
s\omega_0^{2n+1} &=& 0
\eea
where, as usual, $\omega^q_p$ denotes a p-form with ghost number q.

In particular, from Eq.~(\ref{eq:gl}) one has that the nontrivial
solution of the last equation in ~(\ref{eq:tow}) corresponding to
${\cal P}_{2n+2}(F)$ is given by the ghost monomial of degree $(2n+1)$
\be
\omega^{2n+1}_0=Tr\frac{c^{2n+1}}{(2n+1)!}.\label{eq:gmm}
\ee
In what follows we shall assume that the
descent equations ~(\ref{eq:tow}) refer always to the monomials of the
basis ~(\ref{eq:gl}).

\section{Equivalence  with the Russian formula }

In order to solve the descent equations ~(\ref{eq:tow}) we proceed
as in \cite{s} and we introduce  two differential
operators
$\delta$ and ${\cal G}$ defined by
\bea
\delta A=0  \hspace{1.2cm} &,&\hspace{1.2cm}\delta F=0 \nonumber \\
\delta c=-A \hspace{1.2cm} &,&\hspace{1.2cm}\delta \xi=F+A^2,
\label{eq:del}
\eea
and
\bea
{\cal G}A=0 \hspace{1.2cm} &,&\hspace{1.2cm}{\cal G}F=0 \nonumber\\
{\cal G}c=-F \hspace{1.2cm} &,&\hspace{1.2cm} {\cal G}\xi=FA-AF.
\label{eq:ge}
\eea
The operators $\delta$ and ${\cal G}$ are respectively of degree
zero and one and obey the following algebraic relations:
\be
d=-\left[s,\delta\right] \label{eq:s1}
\ee
\be
2{\cal G}=\left[d,\delta\right] \label{eq:s2}
\ee
\be
\left\{ d,{\cal G}\right\}=\left\{ s,{\cal G}\right\}=0 \label{eq:s3}
\ee
\be \left\{ {\cal G},{\cal G} \right\}=\left[{\cal G},\delta \right]=0
 \label{eq:s4}
\ee
 In particular Eq.~(\ref{eq:s1}) shows that the operator $\delta$
 decomposes the exterior derivative $d$ as a BRST commutator.

 Equations (\ref{eq:s1})-(\ref{eq:s4}) define an algebraic setup which,
 as we shall see later, gives a systematic procedure
 for solving the tower~(\ref{eq:tow}).

 To this purpose let us  make use of the following identity
 \be
e^{\delta} s e^{-\delta}=s+d-{\cal G} \label{eq:com}
 \ee
 which is a direct consequence of ~(\ref{eq:s1})-(\ref{eq:s4})
  and of the elementary  formula
 $$
e^{A} B e^{-A}=B+[A,B]+\frac{1}{2!}[A,[A,B]]+
\cdots
$$
 Eq.~(\ref{eq:com}) can be rewritten in a more useful way as
\be
e^{\delta} s=(s+d)e^{\delta}-e^{\delta}{\cal G} \label{eq:com2}
\ee
 Let us apply now Eq.~(\ref{eq:com2}) to the ghost monomial of
 Eq.~(\ref{eq:gmm}). Taking into account that
 $\omega^{2n+1}_0$ belongs to the cohomology of $s$ one gets
\be
(s+d)e^{\delta}\omega_0^{2n+1}-
e^{\delta}{\cal G}\omega_0^{2n+1}=0 \label{eq:t5}
\ee
{}From Eq.~(\ref{eq:s3}) one has
\be
s{\cal G}\omega^{2n+1}_0=-{\cal G} s \omega^{2n+1}_0=0
\ee
which shows that ${\cal G}\omega^{2n+1}_0$ is $s$-invariant.
However, from Eq.~(\ref{eq:ge}) and from the general result~(\ref{eq:gl}),
it follows that ${\cal G}\omega^{2n+1}_0$ cannot
belong to the cohomology of $s$. Hence it is trivial, i.e.
\be
{\cal G}\omega_0^{2n+1}=s\Omega_2^{2n-1} \label{eq:tow3}
\ee
As explained in \cite{s}, this equation yields a chain of
forms $(\Omega^{2n-2}_2,\Omega^{2n-3}_4,\cdots,\Omega^{1}_{2n})$
which obey a tower of descent equations involving
the operators $s$ and ${\cal G}$:
\bea
{\cal G} \Omega_2^{2n-1}&+&s \Omega_4^{2n-3}=0 \nonumber\\
{\cal G} \Omega_4^{2n-3}&+&s \Omega_6^{2n-5}=0 \nonumber\\
{\cal G} \Omega_6^{2n-5}&+&s \Omega_8^{2n-7}=0 \nonumber\\
&\cdots& \nonumber\\
&\cdots& \nonumber\\
{\cal G} \Omega_{2n-2}^3&+&s\Omega^1_{2n}=0
\label{eq:tow2}
\eea
 and
\be
{\cal G} \Omega_{2n}^1=(const.){\cal P}_{2n+2}(F)
\label{eq:pol}
\ee
where ${\cal P}_{2n+2}(F)$ is the invariant monomial of degree
$(2n+2)$.The expression for the $\Omega$-cocycles and for
the constant factor of eq.(\ref{eq:pol}) will be computed
exactly in the next section. Let us focus then, for the time
being, on the proof of the equivalence with the
{\em "Russian formula"}.

Using the algebraic relations (\ref{eq:s1}) and (\ref{eq:s2})
and the ladder (\ref{eq:tow2})-(\ref{eq:pol}) it is easy to
prove that the equation (\ref{eq:t5}) can be iterated to
give
\be
  (s+d)\omega_S+(const.){\cal P}_{2n+2}(F)=0 \label{eq:mr}
\ee
with
\be
\omega_S=e^{\delta}(\omega^{2n+1}_0-\Omega^{2n-1}_2-\cdots-\Omega^{1}_{2n})
\label{eq:mr1}
\ee
Equation~(\ref{eq:mr}) represents our main result. It summarizes in
a unique closed equation the whole solution
of the descent equation (\ref{eq:tow}). Indeed, projecting out from
$\omega_S$ the terms whith a given ghost number one gets:
\be
\omega^{2n+1}_0=Tr\frac{c^{2n+1}}{(2n+1)!},
\label{eq:s11}
\ee
\be
\omega^{2n+1-2p}_{2p}=\frac{\delta^{2p}}{(2p)!}\omega^{2n+1}_0-
\sum_{j=0}^{p-1}\frac{\delta^{2j}}{(2j)!}\Omega^{2n+1-2p+2j}_{2p-2j},
\label{eq:s12}
\ee
for the even space-time form sector and
\be
\omega^{2n}_1=\delta\omega^{2n+1}_0,
\label{eq:s13}
\ee
\be
\omega^{2n-2p}_{2p+1}=\frac{\delta^{2p+1}}{(2p+1)!}\omega^{2n+1}_0-
\sum_{j=0}^{p-1}\frac{\delta^{2j+1}}{(2j+1)!}\Omega^{2n+1-2p+2j}_{2p-2j},
\label{eq:s14}
\ee
for the odd sector and $p=1,2,\cdots,n$.

Equations (\ref{eq:s11})-(\ref{eq:s14})
are nothing but the explicit solution of the ladder (\ref{eq:tow})
obtained in ref.\cite{s}.

It is easy now to compare Eq.(\ref{eq:mr}) with the corresponding
expression given by the {\em"Russian formula"}; this will allow to
establish the cohomological equivalence of the two methods.

To this purpose let us recall that, using the {\em"Russian formula"}
 and the transgression procedure
of \cite{z,zwz,msz,dtv,baulieu,stora}, we can rewrite
 the invariant monomial ${\cal P}_{2n+2}(F)$ as
 \be
 {\cal P}_{2n+2}(F)=(d+s)\tau(A-c,F)
 \label{eq:tf2}
 \ee
 where the Chern-Simons term $\tau(A-c,F)$\footnote{The unusual relative
 minus sign of the combination $(A-c)$ in the
 {\em "Russian formula"} (\ref{eq:tf2}) is due to the convention adopted
 for the BRST transformations (\ref{eq:brst})}
is defined by Eq. (\ref{eq:tf1}),
i.e.
 \be
 {\cal P}_{2n+2}(F)=d\tau(A, F) \label{eq:tf11}
 \ee
 Using Eq.(\ref{eq:tf2}), the expression (\ref{eq:mr}) takes the
 form
 \be
 (s+d)( \omega_S+(const.)\tau(A-c,F))=0
 \ee
 from which, using the triviality of the
 cohomology of $(d+s)$ (see Sect.2) it follows that $\omega_S$
 and $\tau(A-c,F)$ differ only by a trivial $s$ or $d$-coboundary.
 We have thus established the equivalence with the {\em "Russian formula"}.

\section{Solution of the descent equations}

In this section we give a simple procedure for finding the
explicit expressions for the $\Omega$-cocycles of the tower
(\ref{eq:tow2}). We will be able then to give an elegant
and very useful expression for the cocycle $\omega_S$
of Eq.(\ref{eq:mr1}). This will give us a
straightforward way of computing the anomalies for any space-time dimension.

Let us begin by recalling some important
algebraic properties concerning the symmetrized trace of a set
of $n$ matrices $(T_{a_1},T_{a_2},\cdots,T_{a_n})$
belonging to the Lie algebra of G. Following ref.\cite{z,zwz,z2} it is given
by
\be
{\cal
S}Tr(T_{a_1},T_{a_2},\cdots,T_{a_n})=\frac1{n!}\sum_{\pi}Tr(T_{a_{\pi(1)}}
T_{a_{\pi(2)}} \cdots T_{a_{\pi(n)}}) \label{eq:st}
\ee
where the sum is over all permutations $(\pi(1),\pi(2),\cdots,\pi(n))$
of $(1,2,\cdots,n)$. The symmetric invariant polynomial
$ P({\cal F}_1,{\cal F}_2,\cdots,{\cal F}_n)$ in
 the Lie algebra-valued forms $( {\cal F}_1,{\cal F}_2,
\cdots,{\cal F}_n) $ is defined as
 \be
 P({\cal F}_1,{\cal F}_2,\cdots,{\cal F}_n)={\cal F}^{a_1}_1
 \cdots {\cal F}^{a_n}_n STr(T_{a_1},\cdots,T_{a_n})
 \ee
 with ${\cal F}_j={\cal F}^a_j T_a$ $ (j=1,\cdots,n)$.
 The symmetrized trace has the remakable property
 \be
 \sum_{i=1}^{n}{\cal S}Tr(T_{a_1},\cdots,\left[\theta,T_{a_i}\right],
 \cdots,T_{a_n})=0   \label{eq:rp}
 \ee
 with $\theta $ an arbitrary element of the Lie
 algebra.

Now let $\Theta$  and  $(\Lambda_i ,i=1,\cdots,n)$  be  Lie algebra-valued
forms of degree $d_{\Theta}$ and $d_i$.
Then from Eq.~(\ref{eq:rp}), one gets the
useful identity:
\be
\sum_{i=1}^{n}(-1)^{(d_1+\cdots+d_{i-1})d_{\Theta}}P(\Lambda_1,\cdots,
\left[ \Theta,\Lambda_i \right],\cdots,\Lambda_n)=0 \label{eq:ii}
\ee
where the extra sign in each term accounts for the exchange
of $\Theta$ with the forms $(\Lambda_1,\cdots,\Lambda_{i-1})$.

Using the  symmetrized trace we can rewrite the
invariant ghost monomial $\omega_0^{2n+1}(c)$ of Eq.~(\ref{eq:gmm}) as
\be
\omega_0^{2n+1}(c)=\frac1{(2n+1)!}P(c,\underbrace{ c^2,\cdots,c^2}_{n-
\mbox{times}})=\frac1{(2n+1)!}P(c,(c^2)^n) \label{eq:gm}
\ee
where in the last equation we have used Zumino's convention \cite{z,zwz,z2}
$$
P({\cal F}_1,{\cal F}_2,{\cal F}_3,\underbrace{{\cal F},\cdots,{\cal F}}_
{n-\mbox{times}})=P({\cal F}_1,{\cal F}_2,{\cal F}_3,{\cal F}^n)
$$

 We turn now to solve the ladder
  (\ref{eq:tow3})-(\ref{eq:tow2}). Let us start by considering the
  first equation. Taking into account the form of
  $\omega^{2n+1}_0$ and the definition of the differential ${\cal G}$
  of Eq.~(\ref{eq:ge}) one has:
\bea
{\cal G}\omega^{2n+1}_0=\frac1{(2n+1)!}\left[ P({\cal G}c,(c^2)^n)-
nP(c,{\cal G}c^2,(c^2)^{n-1}) \right] \nonumber\\
=\frac1{(2n+1)!}\left[-P(F,(c^2)^n)-nP(c,sF,(c^2)^{n-1}) \right]
\label{eq:f13}
\eea
since ${\cal G}c^2= sF$.

The left hand side of this equation can be
written as an exact $s$-cocycle.
Indeed, the use of Eq.(\ref{eq:ii})
with $\Theta=c, \Lambda_1=c,\Lambda_2=F, (\Lambda_3=\cdots=\Lambda_n=c^2)$
implies
\bea
 &P& (\left[ c,c \right],F,(c^2)^{n-1})
 - P(c,\left[ c,F \right],(c^2)^{n-1})
\nonumber \\
      &-& (n-1)P(c,F,\left[ c,c^2 \right], (c^2)^{n-2})=0
\label{eq:f14}
\eea
or
\be
2P(F,(c^2)^n)+P(sF,c,(c^2)^{n-1})=0
\label{eq:f15}
\ee
due to the Jacobi identity.

{}From
\be
sP(c,F,(c^2)^{n-1})=P(F,(c^2)^n)-P(c,sF,(c^2)^{n-1}).
\label{eq:f16}
\ee
one easily gets, using Eq.( \ref{eq:f15}),
\be
sP(c, F, (c^2)^{n-1})=-\frac{1}{2}P(c,(sF), (c^2)^{n-1})
\label{eq:f17}
\ee
Eq.(\ref{eq:f13}) takes the form
\be
{\cal G}\omega^{2n+1}_0=s \left( \frac1{(2n)!}P(F,c,(c^2)^{n-1}) \right)
\label{eq:f18}
\ee
so that $ \Omega^{2n-1}_2 $ can be identified with
\be
\Omega^{2n-1}_2=\frac1{(2n)!}P(F,c,(c^2)^{n-1}).
\label{eq:f19}
\ee
We proceed now by induction. Let us suppose that the form of
the cocycles $\Omega^{2n-2p+1}_{2p}$ in Eqs.~(\ref{eq:tow2})  is given by
\be
\Omega^{2n-2p+1}_{2p}=\frac{(-1)^{p+1}}{(2n-p+1)!p!}P(F^p,c,(c^2)^{n-p}).
\label{eq:f20}
\ee
For $p=1$ Eq.~(\ref{eq:f20}) reduces just to
 expression ~(\ref{eq:f19}). Acting with ${\cal G}$
 on Eq.(\ref{eq:f20}) one has:
 \bea
{\cal G}\Omega^{2n-2p+1}_{2p} &=& \frac{(-1)^{p+1}}{(2n-p+1)!p!} ({\ }
                     -P(F^{p+1},(c^2)^{n-p}) \nonumber\\
    &-& (n-p)P(F^p,c,sF,(c^2)^{n-p-1}) {\ }).
\label{eq:f21}
\eea
Repeating the same arguments of Eqs.(\ref{eq:f14}),(\ref{eq:f15})
it is not difficult to get the identity
\be
2P(F^{p+1},(c^2)^{n-p})+(p+1)P(sF,F^p,c,(c^2)^{n-p-1})=0
\label{eq:f22}
\ee
which combined with
\bea
sP(F^{p+1},c,(c^2)^{n-p-1}) &=& (p+1)P(sF,F^p,c,(c^2)^{n-p-1}) \nonumber \\
  & +& P(F^p,(c^2)^{n-p})
\label{eq:f23}
\eea
leads to the equation
\be
{\cal G}\Omega^{2n-2p+1}_{2p}=-s\Omega^{2n-2(p+1)+1}_{2(p+1)}
\label{eq:f24}
\ee
with
\be
\Omega^{2n-2(p+1)+1}_{2(p+1)}=\frac{(-1)^{p+2}}{(2n-p)!(p+1)!}
P(F^{p+1},c,(c^2)^{n-p-1}).
\label{eq:f25}
\ee
Therefore we can conclude that the expressions
 for the $\Omega$-cocycles given in Eq.~(\ref{eq:f21}) are indeed solutions
 of the tower~(\ref{eq:tow3})-(\ref{eq:tow2}). In particular, for
 $p=n$ one finds
\be
\Omega^1_{2n}=\frac{(-1)^{n+1}}{(n+1)!n!}P(F^n,c)
\label{eq:f26}
\ee
and Eqs.~(\ref{eq:pol}),~(\ref{eq:mr}) take now a perfect defined form
\be
{\cal G}\Omega^1_{2n}=\frac{(-1)^n}{(n+1)!n!}P(F^{n+1})
\label{eq:f27}
\ee
without any unspecified constant.

Let us close this section with a very important remark. As one can see
from Eq.(\ref{eq:del}) the operator $\delta$ acts as a translation on
the ghost $c$ with an amount given by $(-A)$. Then
if $e^{\delta}$ acts on a quantity which depends only on $A, F$ and
$c$ it has the simple effect of translating $c$. This implies that the
cocycle $\omega_S$ of Eq.(\ref{eq:mr1}) can be written as
\be
\omega_S=\omega^{2n+1}_0(c-A)-\Omega^{2n-1}_2(c-A,F)-\cdots-
\Omega^1_{2n}(c-A,F)
\label{eq:f28}
\ee
which, using Eq.(\ref{eq:f20}), gives:
\be
\omega_S(F,c-A)=\sum_{p=0}^n \frac{(-1)^{p}}{(2n-p+1)!p!}
P(F^p,c-A,\left[ (c-A)^{2} \right]^{n-p}).
\label{eq:f29}
\ee
This remarkable equation collects in a
very elegant and simple form the solution of
the descent equations (\ref{eq:tow}). In particular, as
we will see in next section, we can immediately write down the
expressions for the gauge anomalies for  any space-time dimension.

\section{Some examples}
This section is devoted to use the closed
expression (\ref{eq:f29}) to discuss some explicit examples.

\subsection{The case n=1}
In this case, relevant for the two-dimensional anomaly and for
the three dimensional Chern-Simons term, the $\omega_S$ cocycle reads
\be
\omega_S(F,c-A)=\frac1{3!}P(c-A,(c-A)^2)-\frac1{2!}P(F,c-A)
\label{eq:e1}
\ee
and by expanding in power of $c$ we get
\be
\omega^3_0=\frac1{3!}P(c,c^2)=\frac1{3!}Tr(c^3),
\label{eq:e2}
\ee
\bea
\omega^2_1 &=& -\frac1{3!}P(A,c^2)-\frac1{3!}P(c,\left[c,A\right])
                               \nonumber\\
           &=& -Tr(Ac^2)=Tr(\xi c)-sTr(Ac),
\label{eq:e3}
\eea
\bea
\omega^1_2 &=& \frac1{3!}P(c,A^2)+\frac1{3!}P(A,\left[ c,A \right])
      -\frac1{2!}P(c,F) \nonumber\\
 &=& \frac1{2}\left( Tr(c A^2)-Tr(cF)\right]=-\frac1{2}Tr\left[c(dA)\right)
\label{eq:e4}
\eea
\bea
\omega^0_3 &=& -\frac1{3!}P(A,A^2)+\frac1{2!}P(F,A) \nonumber\\
           &=& \frac1{2}Tr(AdA+\frac2{3}A^3).
\label{eq:e5}
\eea
One can easily recognize that the expresions (\ref{eq:e3})-(\ref{eq:e5})
coincide,
modulo coboundaries, with the solution given by Zumino, Wu and
Zee \cite{zwz}. In particular $\omega^1_2,\/ \omega^2_1$ and
$\omega^0_3 $ give
respectively the two dimensional gauge anomaly, the Schwinger term and the
three dimensional Chern-Simons action.

\subsection{The case n=2}

For n=2 expression ~(\ref{eq:f29}) takes the form
\bea
\omega_S &=&\frac1{5!}P(c-A,\left[(c-A)^2\right]^2) \nonumber\\
         &-& \frac1{4!}P(F,c-A,(c-A)^2)+
              \frac1{3!2!}P(F^2,c-A).
\label{eq:e6}
\eea
In particular $\omega^0_5$ and $\omega^1_4$ are computed to be
\be
\omega^0_5=-\frac{1}{12}Tr(\frac{1}{10}A^5-\frac{1}{2}FA^3+F^2A),
\label{eq:e7}
\ee
\be
 \omega^1_4=   \frac{1}{4!} Tr \left( c
    ( A^4 - F A^2 -A^2 F - A F A + 2 F^2 ) \right)
\label{eq:e8}
\ee
and give respectively the generalized five-dimensional
Chern-Simons term and the four-dimensional gauge anomaly.
Again, they coincide, modulo coboundaries, with that of
ref.\cite{zwz}.

\subsection{The general case}
It is strightforward now to generalize the previous examples and find out
from the formula ~(\ref{eq:f29}) the general form for the cocycles
$\omega^{2n+1-2p}_{2p}$ for any  n and p.
We shall give here only the expressions of the
generalized $(2n+1)$-dimensional Chern-Simons term and of
the $2n$-dimensional anomaly:
\be
\omega^{0}_{2n+1}=  \sum_{p=0}^n\frac{(-1)^{p+1}}{(2n-p+1)!p!}
P(F^p,A,(A^2)^{n-p})
\label{eq:e9}
\ee
and
\bea
\omega^1_{2n } &=&  \sum_{p=0}^{n}  \frac{(-1)^p}{(2n-p+1)!p!}
                    P(c,F^p,(A^2)^{n-p})   \nonumber \\
  &+&    \sum_{p=0}^{n}  \frac{(-1)^p (n-p)}{(2n-p+1)!p!}
                  P([c,A],F^p,A,(A^2)^{n-p-1})
\label{eq:e10}
\eea
Let us conclude by emphasizing that, actually, expression ~(\ref{eq:e10})
represents one of the most closed algebraic formula for the gauge
anomaly in any space-time dimension.

\subsection{Conclusions}
We have proved that the method proposed by \cite{s} for solving the
descent equations associated with the Wess-Zumino consistency
condition is completely equivalent to that based on the well-known
{\em "Russian formula"}\cite{z,msz,dtv,baulieu,stora}.
Moreover, it naturally extends to the case of the
gravitational anomalies as well as to the recently proposed
topological field theories.

\vspace{4mm}

{\bf ACKNOWLEDGMENTS}

We are grateful to all the members of the Institut
f\"{u}r Theoretische
Physik of the Technische Universit\"{a}t Wien for
useful discussions and comments.
On of us (LT) would like to thank Prof. W.Kummer for the
extended hospitality at the Institut.

\end{document}